\documentclass[english,notitlepage,twocolumn]{revtex4-1}
\usepackage[T1]{fontenc}
\usepackage[latin9]{inputenc}
\usepackage{amsmath}
\usepackage{graphicx}

\makeatletter
\def\ro{|\rho_{\rm i}\rangle} 

\usepackage{changepage}

\@ifundefined{showcaptionsetup}{}{%
 \PassOptionsToPackage{caption=false}{subfig}}
\usepackage{subfig}
\makeatother

\usepackage{babel}
\begin{document}

\title{Lattice Model for Production of Gas}

\author{M. Marder, Behzad Eftekhari}
\affiliation{Department of Physics, The University of Texas at Austin,
  USA}
\email{marder@mail.utexas.edu}
\author{Tadeusz W. Patzek}
\affiliation{Petroleum Engineering Research Center, King Abdullah
  University of Science and Technology, Saudi Arabia}
\begin{abstract}
We define a lattice model for rock, absorbers, and gas that makes
it possible to examine the flow of gas to a complicated absorbing
boundary over long periods of time. The motivation is to deduce the
geometry of the boundary from the time history of gas absorption.
We find a solution to this model using Green's function techniques,
and apply the solution to three absorbing networks of increasing complexity. 
\end{abstract}
\maketitle

\paragraph{Introduction.}

Hydrofracturing \cite{Turcotte.14,Marder2016} liberates gas from
mudstones, often called shales, that are brittle residues of seabeds
buried thousands of meters deep and around 30 meters thick. The extraction
process involves creating a fracture network that extends over a distance
on the order of a kilometer in one direction, and on the order of
200 meters in the perpendicular direction. Diffusion in the mudstone
is very slow, but once gas reaches the fractures it travels rapidly
to a wellbore and from there up to the surface. The time history of
gas production can be used to extract information about the dimensions
and structure of the fractured network \cite{Patzek.13,Patzek.14}.

This setting motivates the creation of a simple model for the flow
of gas from an infinite two-dimensional plane into a bounded but geometrically
complex absorbing boundary. The problem bears a family resemblance
to Diffusion-Limited Aggregation
\cite{witten1981diffusion,Sahimi1993,Halsey2000
}
because it involves diffusion on a lattice into a cluster. However
in this problem the cluster is specified at the outset and does not
change. Because the dynamics are linear, the model is analytically
tractable using Green's functions \cite{Economou.83,Berciu.10,Joyce2017}.
The task is to find the time dependence for the gas reaching the absorbers
because this models the process of gas production.

\begin{figure}
\centering{}\includegraphics[width=1\columnwidth]{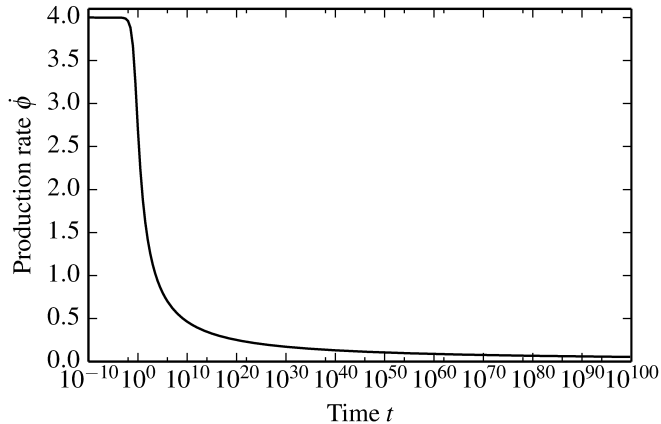}\caption{Production rate of a single absorber over very long time.\label{fig:OneSink}}
\end{figure}

\paragraph{Model Definition.}

We pose the model on a two-dimensional square lattice, whose nodes
are indexed by $j.$ On every node we place a time-dependent density
of gas $\rho_{j}(t).$ We denote the state vector of densities by
$|\rho(t)\rangle,$ and the density at location $j$ by $\rho_{j}(t)=\langle j|\rho(t)\rangle$.
The problem begins at time $t=0$ with a uniform initial density $\ro\equiv|\rho(0)\rangle;\quad\langle j\ro=\rho_{j}(0)=1.$
Every site is either rock through which gas diffuses or else an absorber
that collects all gas reaching it and sends none back. To keep track
of absorbers we define $\theta_{j}$ by 
\begin{equation}
\theta_{j}=\begin{cases}
1 & \text{if \ensuremath{j} is unbroken rock)}\\
0 & \text{if \ensuremath{j} is an absorber) }
\end{cases}.\label{eq:theta}
\end{equation}

Between every pair of nodes $j,j'$ there is a link of strength $k_{jj'}=k_{j'j}$.
We restrict ourselves here to nearest-neighbor coupling so $k_{jj'}$
is 1 if $j$ and $j'$ are nearest neighbors and 0 otherwise. The
time evolution of $\rho$ is given by
\begin{equation}
\frac{\partial}{\partial t}|\rho\left(t\right)\rangle=\hat{H}|\rho\left(t\right)\rangle,\label{eq:dH}
\end{equation}
where
\begin{equation}
\langle j|\hat{H}|j'\rangle=\theta_{j}\left[k_{jj'}-B\delta_{jj'}\right]\theta_{j'};\quad B\equiv\sum_{j'}k_{jj'}=4.\label{eq:ki}
\end{equation}
 The $\theta$ functions ensure that when gas reaches an absorber
it disappears. Note that $\hat{H}$ is real and symmetric, with all
off-diagonal entries either 0 or 1. The Hamiltonian $\hat{H}$ has
as many zero eigenvalues as there are absorbers. If however one projects
down onto the space of all sites $j$ where $\theta_{j}=1,$ then
these zero eigenvalues disappear. In this space $\hat{H}$ has a complete
set of eigenvectors $|\alpha\rangle$ with eigenvalues $E_{\alpha}$.
This observation is not computationally useful when the spectrum is
continuous. However if one has a closed region completely surrounded
by absorbers, with a finite number $N_{s}$ of interior lattice points,
then the spectrum is discrete, eigenvalues and eigenvectors can be
computed explicitly, and the rate at which gas is produced is 
\begin{equation}
\dot{\phi}_{{\rm interior}}=-\sum_{\alpha=1}^{N_{s}}E_{\alpha}e^{E_{\alpha}t}|\langle\alpha\ro|^{2}.\label{eq:phi0}
\end{equation}

\paragraph{Green's function solution for continuous spectrum.}

\begin{figure}[t]
\subfloat[\label{fig:Network-following-the}Network exhibiting the schematic
geometry for hydrofractured horizontal well from \cite{Patzek.13}. ]{\includegraphics[width=1\columnwidth]{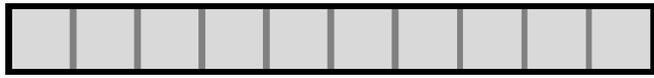}}\\
\subfloat[Spectral function from Eq. (\ref{eq:phidot2}) for network of Figure
(\ref{fig:Network-following-the}).\label{fig:Spectral-function-from}]{\includegraphics[width=1\columnwidth]{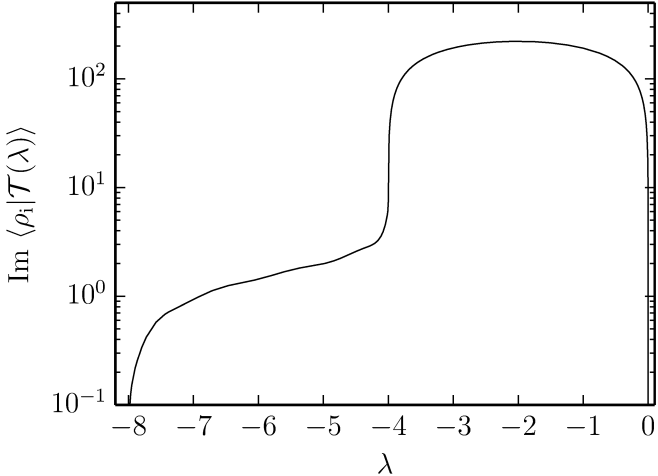}}\\
\subfloat[Cumulative gas production for network of Figure (\ref{fig:Network-following-the})
(color online). The lower (red) curve gives the production from the
interior only.\label{fig:Cumulative-gas-production}]{\includegraphics[width=1\columnwidth]{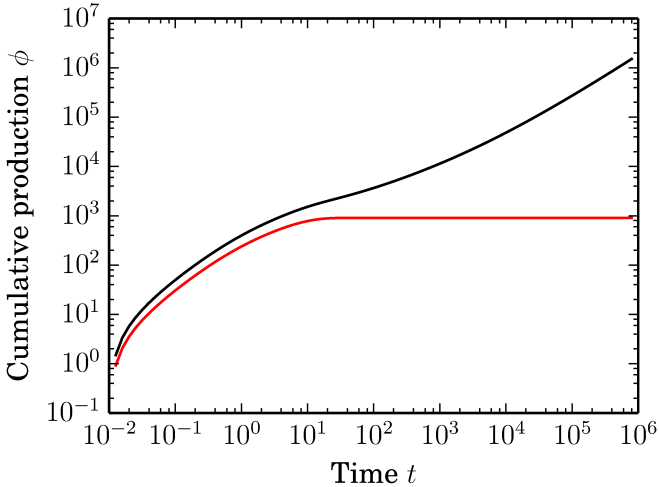}}\caption{Gas production from assembly of absorbers that mimics a simplified
geometry for hydrofractured wells.\label{fig:Gas-production-from-1}}
\end{figure}

\begin{figure}
\includegraphics[width=1\columnwidth]{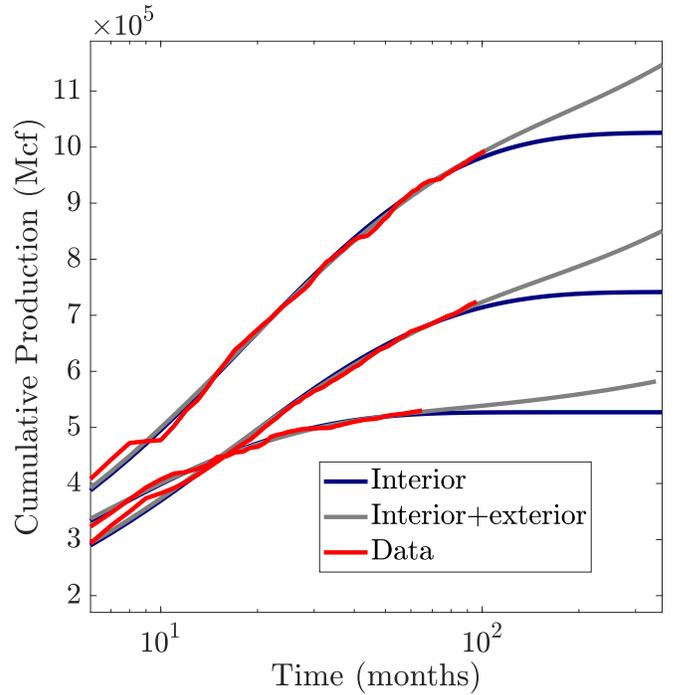}

\caption{Three wells from the Barnett shale that have produced long enough
that the contribution from the exterior region is starting to become
visible. Mcf=1000 cubic feet. (Color online). \label{fig:Three-wells-from}}
\end{figure}

The exterior region of the absorbing network has an infinite number
of sites, the spectrum has a continuous component, and explicit calculation
of all eigenvalues is impossible. Obtaining the continuous spectrum
requires a different approach. Define the Green function \cite{Economou.83}
$\hat{G}(E)=\left(E-\hat{H}\right)^{-1}$and the unperturbed Green
function $\hat{G}_{0}(E)=\left(E-\hat{H}_{0}\right)^{-1}$ for the
lattice without any absorbers. The full Hamiltonian is $\hat{H}=\hat{H}_{0}+\hat{H}_{1}$
where
\begin{equation}
\hat{H}_{1}=-\sum_{jj'}|j\rangle\left(k_{jj'}-B\delta_{jj'}\right)\left(1-\theta_{j}\theta_{j'}\right)\langle j'|.\label{eq:H1-b}
\end{equation}

Note that $\hat{H}_{1}$ is nonzero only in the space spanned by absorbers
and their nearest neighbors. One can obtain the continous spectrum
from computations in this finite-dimensional space. The full Green
function can be obtained from the scattering $\hat{T}$ matrix defined
by $\hat{G}=\hat{G}_{0}+\hat{G}_{0}\hat{T}\hat{G}_{0}.$ To find gas
production it turns out one does not need the complete $\hat{T}$
matrix, but just the state vector produced by its action on the initial
state $|{\cal T}(E)\rangle\equiv\hat{T}(E)|\rho_{{\rm i}}\rangle.$
This state vector is determined by the linear system
\begin{equation}
\left(1-\hat{H}_{1}\hat{G}_{0}(E)\right)|{\cal T}(E)\rangle=\hat{H}_{1}\ro\label{eq:Teqn4}
\end{equation}
and then gas production due to the continous spectrum of the exterior
is provided by

\begin{align}
\dot{\phi}_{{\rm exterior}} & =-\int_{-8}^{0}\frac{d\lambda}{\pi\lambda}e^{\lambda t}\,\textrm{Im}\lim_{\eta\rightarrow0}\langle\rho_{{\rm i}}|{\cal T}(\lambda-i\eta)\rangle.\label{eq:phidot2}
\end{align}
Because of the limit $\eta\rightarrow0,$ which also applies implicitly
to subsequent equations, Eq. (\ref{eq:phidot2}) includes no contributions
from the gas at interior sites captured by Eq. (\ref{eq:phi0}). The
spectrum of the interior sites is discrete, consists of a finite sum
of delta functions, and these vanish in the limiting process. The
exterior problem can also have a discrete spectrum\cite{sutherland1986localization},
missing from this expression, as we discuss later.

Computation of Eq. (\ref{eq:Teqn4}) requires computing matrix elements
$\lim_{\eta\rightarrow0}\langle j|\hat{G}_{0}(\lambda-i\eta)|j'\rangle$
of the unperturbed lattice Green function. This can be done using
recursion relations due to Morita \cite{Morita.75}, and described
in compact form by Berciu \cite{Berciu.10}. These recursion relations
are exponentially unstable, and the instability is particularly severe
as $\lambda$ approaches the band edge at $\lambda=0$. One must get
very close to the band edge since values as small as $\lambda\approx-10^{-1000}$
are needed for the integral in Eq. (\ref{eq:phidot2}) to converge.
The solution we have adopted is to employ Morita's recursions with
high-precision arithmetic. For absorber networks extending across
100 lattice sites, the recursions require from 100 to 5000 places
of precision. Once the Green function matrix elements have been obtained,
the rest of the computation can be carried out in ordinary double
precision.

\begin{figure}
\subfloat[Network of absorbers created by 550 horizontal and vertical cracks.
\label{fig:Network-of-sinks}]{\includegraphics[width=0.7\columnwidth]{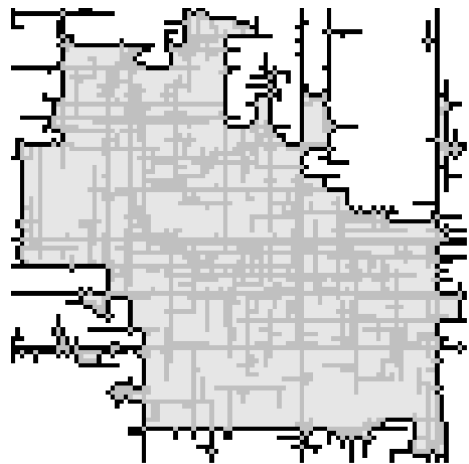}} \\
\subfloat[Continuous spectrum for network in part (a). The sharp peaks come
from small features on the exterior of the network shielded by extended
arms. \label{fig:Continuous-spectrum-for}]{\includegraphics[width=.9\columnwidth]{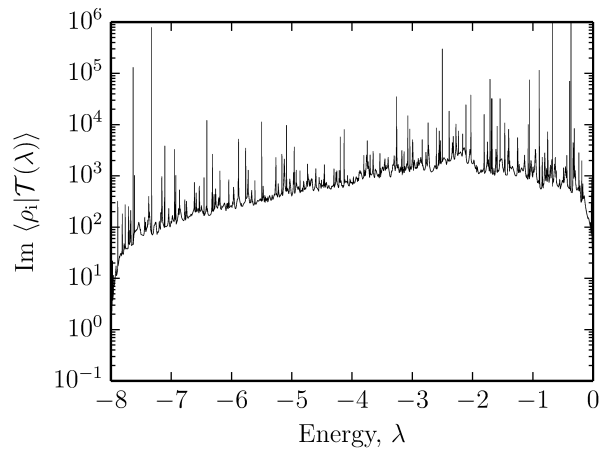}

}\\
\subfloat[Total gas production from network in part (a). \label{fig:Gas-production-from}]{\includegraphics[width=.9\columnwidth]{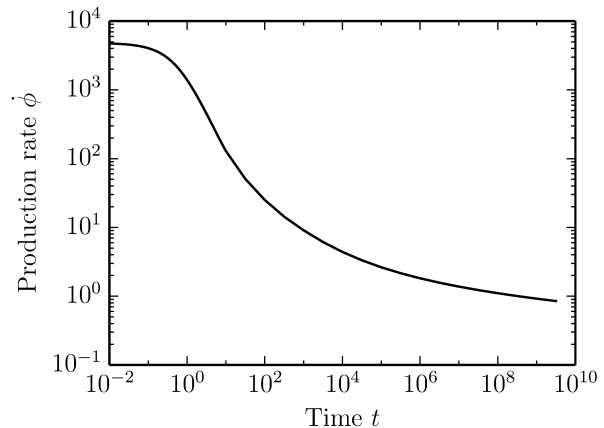}

}\caption{Gas absorption into network of criss-crossing fractures. \label{fig:Gas-absorption-produced}}
\end{figure}

\paragraph{One absorber.}

As a first application of the formalism, consider the case of a single
absorber in an infinite square lattice. The vector $\langle j|{\cal T\rangle}$
has five components. The first corresponds to the absorber at 0, and
the remaining four correspond to the four neighbors of the absorber.
Only $\langle0|{\cal T}\rangle$ has nonzero imaginary part; the remaining
values $\langle j|{\cal T}\rangle=-1,$ where $j=1\dots4,$ and
\begin{equation}
\langle0|{\cal T}(\lambda)\rangle=\lambda+4-\frac{1}{\langle0|\hat{G_{0}}(\lambda)|0\rangle}.\label{eq:tt0}
\end{equation}
This leads to total gas production $\dot{\phi}=\dot{\phi}_{{\rm exterior}}$
\begin{equation}
\dot{\phi}=\int_{-\ln8}^{\infty}\frac{df}{\pi}e^{-te^{-f}}\text{Im}\frac{-1}{\langle0|\hat{G}_{0}(-e^{-f})|0\rangle}\label{eq:OneSinkFinal}
\end{equation}
which for large values of $t$ equals approximately
\begin{align}
\dot{\phi} & \approx\int_{-\ln8}^{\infty}\frac{df}{\pi}\,e^{-te^{-f}}\frac{1}{\tau_{0}+\tau_{1}f+\tau_{2}f^{2}};\label{eq:asymp}\\
\text{where}\  & \tau_{0}=\frac{\left(\ln32\right)^{2}}{4\pi^{2}}+\frac{1}{4},\ \tau_{1}=\frac{\ln32}{2\pi^{2}},\ \tau_{2}=\frac{1}{4\pi^{2}}.\nonumber 
\end{align}
A plot of Eq. (\ref{eq:OneSinkFinal}) appears in Figure (\ref{fig:OneSink}).

\paragraph{Absorbing network with ten stages.}

As a second example, consider the network pictured in Figure (\ref{fig:Network-following-the}).
It is 100 units wide, 10 units high, and has 10 internal regions of
equal size. The interior problem is of size $810\times810.$ By counting
up the number of absorbing faces exposed to gas, one finds that initial
production from the interior must be $\dot{\phi}_{{\rm interior}}(0)=360.$ 

The exterior problem leads to a linear system of equations of size
$323\times323.$ Again counting the number of absorbing faces exposed
to gas on the outside shows that initial production from the exterior
is $\dot{\phi}_{{\rm exterior}}(0)=224$. The same integer comes from
Eq. (\ref{eq:phidot2}) when $t=0;$ this sum rule provides a good
check on the accuracy of the computations and shows that the exterior
problem has no discrete spectrum in this case. The spectral function
$\langle\rho_{{\rm i}}|{\cal {T}}(\lambda)\rangle$ appears in Figure
(\ref{fig:Spectral-function-from}). The interior and total production
appear in Figure (\ref{fig:Cumulative-gas-production}). The well
geometry of Figure (\ref{fig:Network-following-the}) is precisely
the type of geometry used in Ref \cite{Patzek.13} to fit production
of thousands of wells. However, only production from the interior
of the well was previously considered. Here we see that on sufficiently
long time scales, gas arriving from the exterior region leads to a
characteristic upturn on a log-log plot. Analysis of well production
data to extract this characteristic long-time signature is underway.
From the analysis it appears that if one considers a gas-producing
well over the course of thirty years, the extra gas produced by exterior
flow will contribute extra production on the order of 20\%. Figure
(\ref{fig:Three-wells-from}) displays production data from three
wells for which the long-time behavior including production from the
exterior region is just starting to become visible at the ten year
mark; these are some of the oldest existing wells.

\paragraph{Complex fracture network.}

The essential behavior of the examples presented so far could be examined
without much difficulty from a continuum perspective \cite{Carslaw.59}
or using finite element programs such as COMSOL. For a final example,
depicted in Figure (\ref{fig:Network-of-sinks}) we present a geometrically
complex structure produced by 550 intersecting horizontal and vertical
cracks with a power-law length distribution motivated by geophysical
data \cite{Eftekhari.16}. The probability that a fracture have length
$l$ is proportional to $l^{-2.2}$. The interior portion, shown with
grayscale in the figure has 3080 sites in light gray that are rock,
interspersed by interior absorbers in darker gray. Absorbers forming
the exterior of the network are colored in black. The 3080 eigenvalues
and eigenvectors of the interior problem needed for Eq. (\ref{eq:phi0})
can be computed in seconds on a single processor. The sum rule $\dot{\phi}_{{\rm interior}}(0)=3264$
provides a check on the accuracy of these computations. The integrand
of Eq. (\ref{eq:phidot2}) has hundreds of narrow spikes that make
its computation more difficult although still tractable. Counting
up the number of exterior faces on the absorbers gives the sum rule
$\dot{\phi}_{{\rm exterior}}(0)=1556$, but the integral of the continuous
spectrum gives only 1546.53. This is because the external problem
has localized eigenfunctions precisely at $\lambda=-4$. Their weight
can be determined by taking $\eta=10^{-8}$ in Eq. (\ref{eq:phidot2})
and performing the integral in a small neighborhood of $\lambda=-4.$
The contribution from the exterior localized eigenfunctions is 9.47,
and adding this to the integral of the continuous spectrum finally
exhausts the sum rule. 

For large $f=\ln\left(-\lambda\right)$, the continuous spectrum assumes
the asymptotic form of Eq. (\ref{eq:asymp}) with $\tau_{2}=.02533,\quad\tau_{1}=-0.3947,\quad\tau_{0}=1.7877.$
This form is all one needs for the very long time behavior. The continuous
spectrum appears in Figure (\ref{fig:Continuous-spectrum-for}) and
the production rate as a function of time, summing contributions from
the discrete and continuous spectra, appears in Figure (\ref{fig:Gas-production-from}). 

We anticipate that the ability to find precise solutions for the production
history of complex networks of absorbers will assist in studying the
relationship between production data and the complex geometries of
real fractured networks.
\begin{acknowledgments}
BE acknowledges funding from the King Abdullah University of Science
and Technology. We thank David DiCarlo, Carlos Torres-Verdin, and
Larry Lake for useful comments as the work was progressing. Qian Niu
helpfully pointed out the possibility of localized modes in the continuum.
\end{acknowledgments}

\bibliographystyle{apsrev4-1}
\bibliography{/Users/michaelmarder/Dropbox/Shale/Hydrofracture}

\end{document}